\documentclass[12pt,preprint,numberedappendix]{emulateapj}

\usepackage{subeqnarray}
\usepackage{amsmath}
\usepackage{hyperref}
\usepackage{ulem}
\usepackage{stmaryrd}
\hypersetup{colorlinks=true,linkcolor=black,citecolor=blue,urlcolor=black}
\usepackage{natbib}
\def\tt{\bf   }

\newcommand{\Sec}[1]{Section~\ref{#1}}

\newcommand{\Fig}[1]{Figure~\ref{#1}}

\newcommand {\exocam} {{\ttfamily ExoCAM}}

\setcitestyle{citesep={,}}
\begin{document}

\slugcomment{Accepted at ApJL}

\shorttitle{Effects of clouds on water vapor transmission features of terrestrial exoplanets}
\shortauthors{Komacek, Fauchez, Wolf, \& Abbot}

\title{Clouds will likely prevent the detection of water vapor in JWST transmission spectra of terrestrial exoplanets}
\author{Thaddeus D. Komacek$^{1}$, Thomas J. Fauchez$^{2,3,4}$, Eric T. Wolf$^{5,6}$, and Dorian S. Abbot$^{1}$} \affil{$^{1}$Department of the Geophysical Sciences, The University of Chicago, Chicago, IL, 60637 \\ $^{2}$NASA Goddard Space Flight Center, Greenbelt, MD, 20771 \\
$^3$Goddard Earth Sciences Technology and Research (GESTAR), Universities Space Research Association, Columbia, Maryland, USA \\ $^{4}$GSFC Sellers Exoplanet Environments Collaboration \\ $^{5}$Laboratory for Atmospheric and Space Physics, Department of Atmospheric and Oceanic Sciences, University of Colorado, Boulder, CO, 80309 \\ $^{6}$NASA Astrobiology Institute's Virtual Planetary Laboratory, P.O. Box 351580, Seattle, WA 98195, USA \\
%$^2$51 Pegasi b Fellow \\
\url{tkomacek@uchicago.edu}} 
\begin{abstract}
We are on the verge of characterizing the atmospheres of terrestrial exoplanets in the habitable zones of M dwarf stars. Due to their large planet-to-star radius ratios and higher frequency of transits, terrestrial exoplanets orbiting M dwarf stars are favorable for transmission spectroscopy. In this work, we quantify the effect that water clouds have on the amplitude of water vapor transmission spectral features of terrestrial exoplanets orbiting M dwarf stars. To do so, we make synthetic transmission spectra from general circulation model (GCM) experiments of tidally locked planets. We improve upon previous work by considering how varying a broad range of planetary parameters affects transmission spectra. We find that clouds lead to a 10-100 times increase in the number of transits required to detect water features with the James Webb Space Telescope (JWST) with varying rotation period, incident stellar flux, surface pressure, planetary radius, and surface gravity. 
%For the most favorable Earth-like planet, $63$ transits are needed to detect water vapor, which is comparable to the total number of transits observable over the lifetime of JWST. 
We also find that there is a strong increase in the dayside cloud coverage in our GCM simulations with rotation periods $\gtrsim 12 \ \mathrm{days}$ for planets with Earth's radius. This increase in cloud coverage leads to even stronger muting of spectral features for slowly rotating exoplanets orbiting M dwarf stars. 
%We find that even when we include clouds, large amounts of background N$_2$ may be detectable through N$_2$-N$_2$ collision-induced absorption for atmospheres with surface pressures $\gtrsim 4 \ \mathrm{bars}$. 
We predict that it will be extremely challenging to detect water transmission features in the atmospheres of terrestrial exoplanets in the habitable zone of M dwarf stars with JWST. However, species that are well-mixed above the cloud deck (e.g., CO$_2$ and CH$_4$) may still be detectable on these planets with JWST.
%lim = 250 words
\end{abstract}
\section{Introduction}
\indent The upcoming launch of JWST and the future space mission concepts LUVOIR/HabEx/OST promise the characterization of terrestrial exoplanet atmospheres. Previous one-dimensional simulations, which cannot properly account for clouds, have indicated that JWST will be able to observe the atmospheres of potentially habitable exoplanets orbiting M dwarf stars and detect molecular signatures of life in these atmospheres \citep{Barstow:2016aa,Morley:2017aa,Lincowski:2018aa,Lincowski:2019aa,Lustig-Yaeger:2019aa}. However, clouds and hazes have affected observations of exoplanet atmospheres with the Hubble and Spitzer space telescopes by muting signatures of molecular features in transmission \citep{Kreidberg2014b,Sing2015,Crossfield:2017aa}. If clouds or hazes are present at the planetary limb, they pose a problem for transmission spectra of terrestrial exoplanets because of long path lengths through the atmosphere \citep{Moran:2018aa,Badhan:2019aa,Fauchez:2019aa,Lustig-Yaeger:2019aa}. \\
\indent Surface liquid water is considered a necessary constituent of a habitable world \citep{Kasting:1993aa}, and ideally we would like to detect water vapor spectral signatures as an indicator of the habitability of a terrestrial exoplanet. Given the narrow thermodynamic range of liquid water stability, and typical lapse rates in planetary atmospheres, any planet with abundant liquid surface water will also have clouds condensing in its atmosphere. This suggests that hunting for water spectral signatures will be confounded by clouds in terrestrial planet atmospheres. On tidally locked planets with hot daysides and cold nightsides, upwelling on the dayside carries moist air to low pressures. This moist air condenses as it is lifted, leading to strong dayside cloud cover on tidally locked terrestrial planets that have surface water \citep{Yang:2013,kopparapu2017,Haqq2018,Fauchez:2019aa,Komacek:2019aa,Suissa:2019aa,Yang:2019aa}. If this dayside cloud cover extends to the terminator, it could significantly hinder the detection of molecular features in transmission. \\
%A wide range of previous work has shown that the atmospheric circulation of slowly rotating and tidally locked terrestrial exoplanets orbiting M dwarf stars is significantly different than that for Earth-like planets orbiting Sun-like stars \citep{Joshi:1997,Merlis:2010,Yang:2013,Hu:2014aa,Yang:2014,Carone:2015aa,Charnay:2015,Koll:2016,Kopp:2016,Turbet:2016aa,Fujii:2017aa,kopparapu2017,Noda:2017aa,Wolf:2017aa,Haqq2018,Lewis:2018aa,way:2018,Komacek:2019aa,Wolf:2019aa,Yang:2019aa}. 
\indent Recent climate modeling has begun to explore how clouds affect the detection of transmission spectral features with JWST. Using the one-dimensional climate and photochemical models of \cite{Lincowski:2018aa}, \cite{Lustig-Yaeger:2019aa} found that clouds inhibit the detection of water features on TRAPPIST-1e. However, three-dimensional simulations are necessary to accurately simulate cloud and water vapor mixing ratios. By post-processing three-dimensional GCM experiments, \cite{Fauchez:2019aa} and \cite{Suissa:2019aa} found that water vapor is challenging to detect in the atmospheres of terrestrial planets in the habitable zone due to the presence of clouds. However, \cite{Fauchez:2019aa} analyzed simulations only varying the atmospheric composition for individual planets in the TRAPPIST-1 system, and \cite{Suissa:2019aa} considered only the joint effects of varying incident stellar flux and rotation period. \\
\indent In this work, we consider how a much broader range of possible planetary parameters affects transmission spectra. To do so, we post-process the 3D GCM output of \cite{Komacek:2019aa} to make simulated JWST observations of transmission spectra for planets orbiting late-type M dwarf stars. We find that clouds make water vapor transmission spectral features challenging to detect with JWST over a wide range of planetary parameters. 
%To simulate the effect of water clouds on transmission spectra, we make synthetic transmission spectra using the Planetary Spectrum Generator \citep{Villanueva:2018aa} from the simulations of \cite{Komacek:2019aa} with the \exocam \ GCM \citep{Wolf:2015,Kopp:2016,kopparapu2017,Wolf:2017,Wolf:2017aa}. 
We study the difference in transmission spectra when including and not including clouds, along with the effects of varying rotation rate, incident stellar flux, surface pressure, planetary radius, surface gravity, and cloud particle size. 
% Stress that this is just to show how planetary parameters affect spectra and we're not looking at details of different absorbers for an Earth-like atmosphere.
In \Sec{sec:methods}, we describe our GCM experiments and how we post-process our GCM results to simulate transmission spectra. We show how clouds and varying planetary parameters impact transmission spectra in \Sec{sec:results}, along with estimating the number of transits needed to detect water vapor transmission features with JWST. We discuss our results and conclude in \Sec{sec:disc}.
%\clearpage
\section{Methods}
\label{sec:methods}
\subsection{GCM setup}
To simulate the atmospheres of tidally locked terrestrial exoplanets, we use the \exocam \ GCM\footnote{\url{https://github.com/storyofthewolf/ExoCAM}} \citep{Wolf:2015}. \exocam \ is a version of the Community Atmosphere Model version 4 with updated correlated-k radiative transfer and water vapor continuum absorption, with spectral coefficients from HITRAN 2012. \exocam \ has been used in a wide range of studies of the atmospheres of terrestrial exoplanets \citep{Kopp:2016,kopparapu2017,Wolf:2017,Wolf:2017aa,Haqq2018,Komacek:2019aa,Yang:2019aa}. We vary the rotation period, surface pressure, incident stellar flux, planetary radius, surface gravity, and cloud particle size over a wide range relevant for terrestrial exoplanets. The first column of Table \ref{table:transit} shows our considered variations in planetary parameters. Specifically, we use the GCM results for planets orbiting a late-type M dwarf star with $T_\mathrm{eff} = 2600 \ \mathrm{K}$ from \cite{Komacek:2019aa} (see their Table 3). We conduct additional simulations to cover a range of dynamical regimes, including fast, intermediate, and slow rotators. If it is not explicitly stated in Table \ref{table:transit} that a parameter is varied, we keep its value fixed to that of Earth. As a result, this suite of GCM experiments varies planetary parameters individually, and includes some combinations of rotation period and incident stellar flux that are inconsistent with Kepler's laws. We analyze these simulations in order to examine how each of these factors individually affect transmission spectra. \\
\indent We consider an atmosphere consisting of only N$_2$ and H$_2$O on a tidally locked aquaplanet with a 50 m deep slab ocean and zero obliquity. As a result, in this work we focus on how clouds and varying planetary parameters affect transmission spectra. We conducted additional GCM experiments including Earth-like abundances of CO$_2$ and CH$_4$ in order to test the sensitivity of our results to additional greenhouse gases. To determine if the parameterized ice cloud particle size in \exocam \ affects our results, we also include a suite of sensitivity tests with varying ice cloud particle size. Our range of considered ice cloud particle size is $20-200~\mu\mathrm{m}$ (see Table \ref{table:transitcld}), chosen to cover the range of ice cloud particle size in the parameterization of \cite{Rasch:1998} used in \exocam. %We do not include ocean dynamics in our simulations, which has been shown to affect the climate of tidally locked terrestrial exoplanets \citep{Hu:2014aa,Yang:2019ab}. 
All simulations use a horizontal resolution of $4^\circ \times 5^\circ$ with 40 vertical levels and a timestep of 30 minutes. The GCM results presented in this work are averaged over the last 10 years of simulation time.
\subsection{Simulated observables}
To simulate transmission spectra from our GCM output, we use the Planetary Spectrum Generator\footnote{\url{https://psg.gsfc.nasa.gov}} (PSG, \citealp{Villanueva:2018aa}). We use the moderate spectral resolution mode of PSG, which employs the correlated-k technique for radiative transfer, while multiple scattering from aerosols is performed using the discrete ordinates method. The molecular spectroscopy is based on the HITRAN 2016 database \citep{Gordon:2017}, which is complemented by UV/optical data from the MPI database \citep{Keller-Rudek:2013aa}.
%to obtain line lists of water and N$_2$-N$_2$ collision-induced absorption and optical properties of water clouds.} 
We take the temperature, molecular abundance, and liquid and ice cloud profiles from the GCM at each latitude point on the limb as input for PSG. We then use PSG to make a simulated transmission spectrum at every GCM grid point along the terminator, each of which comprises $4^\circ$ of latitude. To make the planetary transmission spectrum, we average the spectra over all latitudinal grid points along the terminator with equal weighting of each spectrum, the same method as used in \cite{Fauchez:2019aa} and \cite{Suissa:2019aa}. Note that this method only takes into account transmission through the limb, but transmission through the cloudier dayside may further reduce the amplitude of transmission spectral features \citep{Caldas:2019aa}. \\
\indent We simulate the transmission spectra for $R = 300$ from $0.6-5.3 \ \mu\mathrm{m}$ relevant for the NIRSpec/PRISM instrument on JWST, which has been shown to be the ideal instrument for JWST characterization of terrestrial exoplanets \citep{Batalha:2018aa,Fauchez:2019aa,Lincowski:2019aa,Lustig-Yaeger:2019aa}. We use the PSG imager noise model and do not include a noise floor in our simulated spectra. As a result, our results can be considered lower limits on the number of required transits to detect water vapor. Note that \cite{Suissa:2019aa} showed that including a noise floor greatly affects the detectability of water features, finding that water vapor is not detectable if the JWST noise floor is $\ge 5~\mathrm{ppm}$. 
%, but note that the noise floor of NIRSpec is expected to be $\sim 20-30 \ \mathrm{ppm}$ \citep{Greene:2015}. 
\section{The dependence of transmission spectra on planetary parameters}
\label{sec:results}
\subsection{Transmission spectra with and without clouds}
\label{sec:cld}
\begin{figure}
\centering
\includegraphics[width=0.5\textwidth]{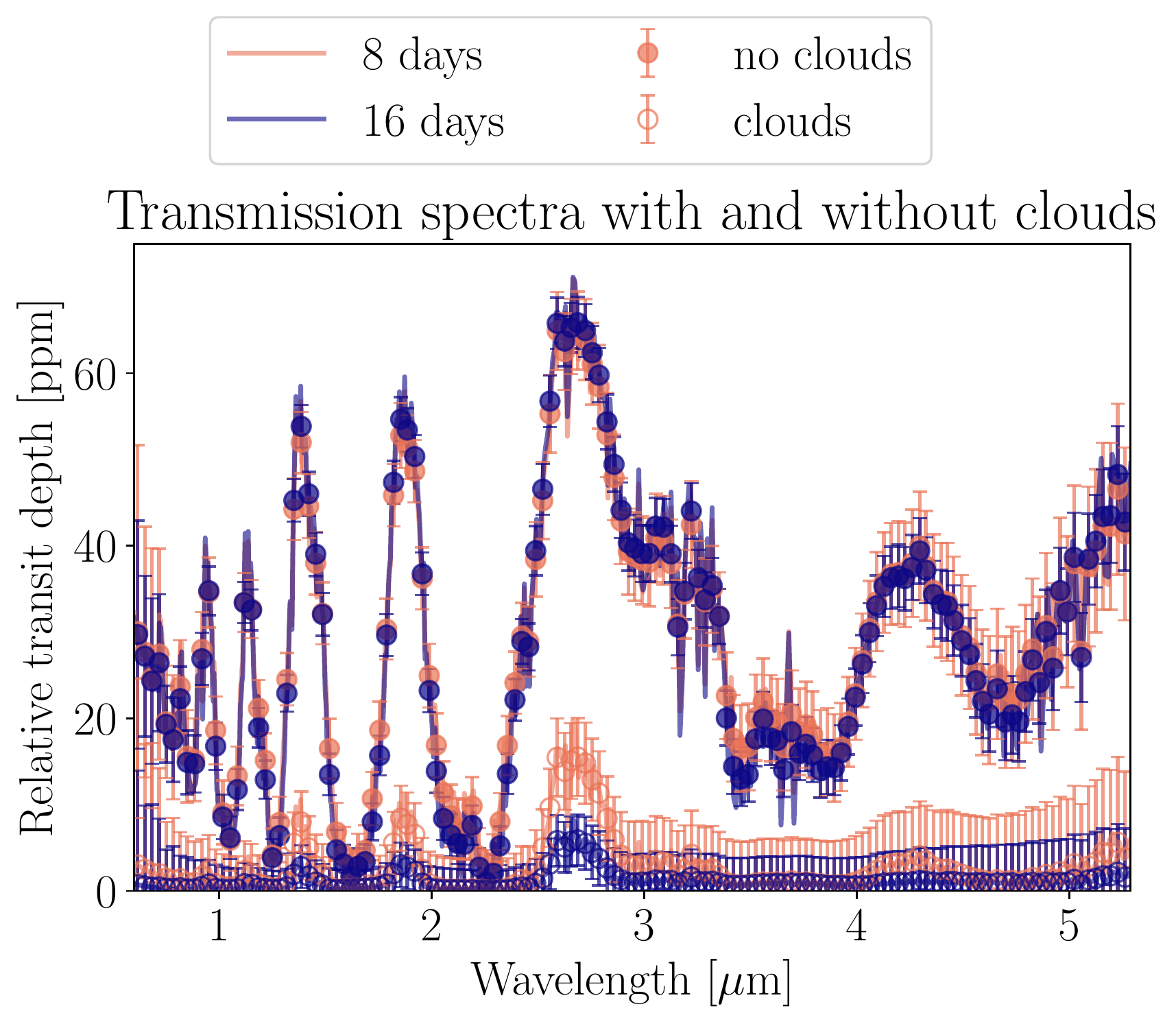}
\caption{ \textbf{Clouds truncate spectral features.} Simulated transmission spectra ignoring clouds (filled points) and including clouds (empty points) for planets with a rotation period of 8 days (orange) and 16 days (blue) orbiting a late-type M dwarf star. 
%Spectra are made by post-processing the \exocam \ GCM with the Planetary Spectrum Generator. 
Lines show simulated spectra from 30 transits with JWST NIRSpec/PRISM, while points with $1\sigma$ errorbars show simulated observations binned to $R = 30$. A transit depth of zero corresponds to the continuum level of the planetary transmission, and the relative transit depth corresponds to the depth of features relative to the continuum. All of the absorption features are due to water except for the $4.1~\mu\mathrm{m}$ feature, which is due to N$_2$-N$_2$ collision-induced absorption. When we include clouds, transmission spectral features of water are significantly diminished. Transmission features are especially weak for planets with long rotation periods.}
\label{fig:cldvsnocld}
\end{figure}
\begin{figure}
\centering
\includegraphics[width=0.5\textwidth]{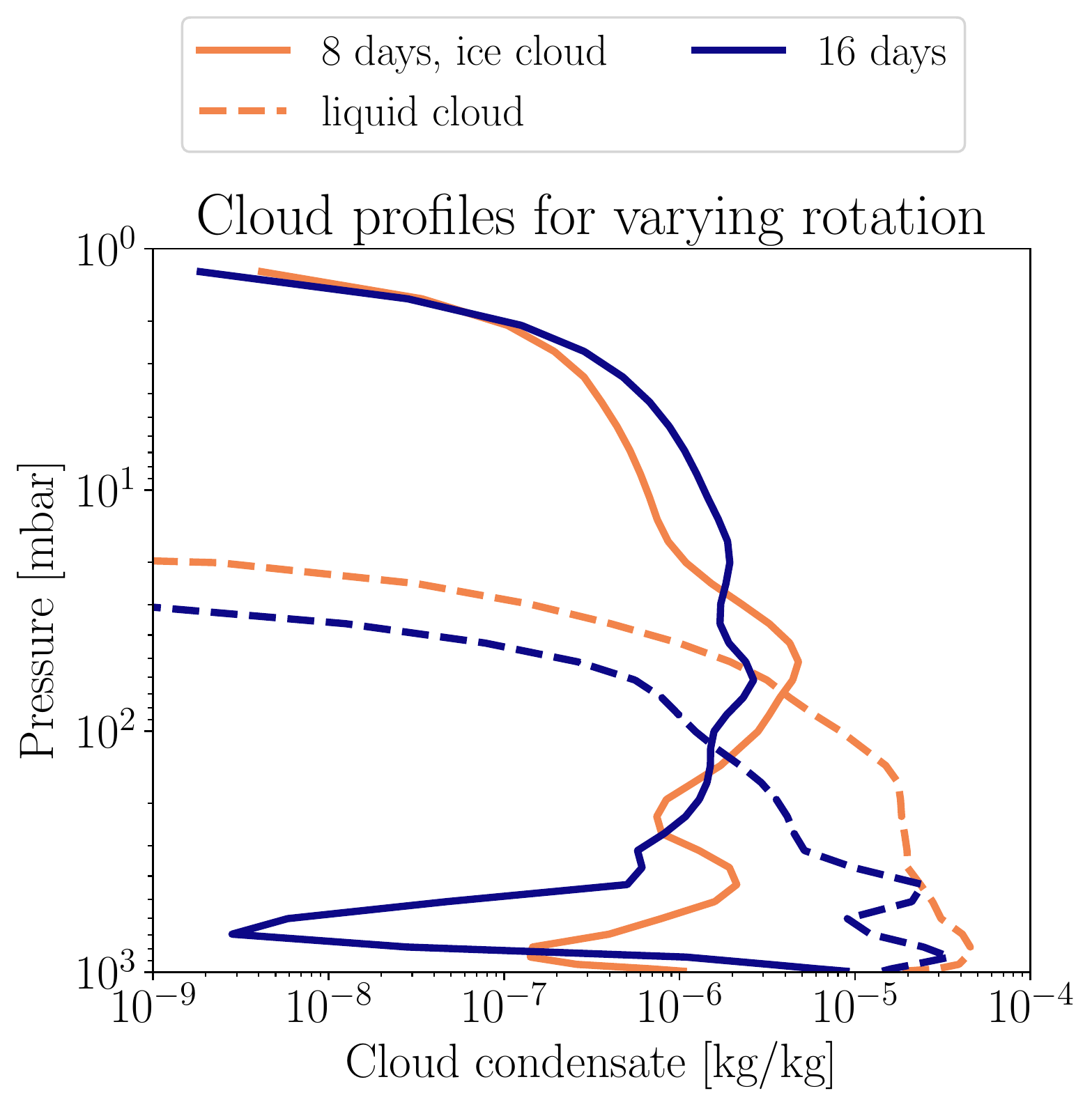}
\caption{\textbf{Rotation can greatly impact cloud coverage.} Ice cloud condensate (sold lines) and liquid cloud condensate (dashed lines) from \exocam \ simulations of a planet with a rotation period of 8 days (orange) and 16 days (blue) orbiting a late-type M dwarf star, with all other parameters fixed to that of Earth. Profiles are averaged along the terminator. We find an increase in the high ice cloud cover between rotation periods of 8 and 16 days, leading to weaker water vapor absorption features.}
\label{fig:cldtrans}
\end{figure}
Our simulated transmission spectra depend strongly on whether we include clouds. \Fig{fig:cldvsnocld} shows simulated transmission spectra from 30 transits with JWST NIRSpec/PRISM. We show results from two GCM experiments with rotation periods of 8 and 16 days, simulating the transmission spectra both including and not including the effects of clouds. We find that when we do not include the effects of clouds in our simulated transmission spectra, transmission spectral features are deep and well above the level of the noise. However, when we include the effects of clouds, the transmission spectral features are strongly muted, with a maximum depth of $\sim 20 \ \mathrm{ppm}$ that is comparable to the expected noise floor of JWST NIRSpec \citep{Greene:2015}. \\
%This depth is similar to the noise floor of JWST NIRSpec.  \\
\indent Cloud muting of transmission spectral features is particularly strong for slowly rotating planets. In \Fig{fig:cldvsnocld}, spectral features for the case with a rotation period of 16 days have an amplitude less than half that of the 8 day rotation period case. \Fig{fig:cldtrans} shows the liquid and ice cloud condensate at the terminator for the 8 and 16 day rotation period cases. As in \cite{Haqq2018}, we find in that there is a greater amount of high-altitude ice cloud condensate in our more slowly rotating simulations. The increase in high-altitude ice cloud condensate leads to the more strongly muted spectral features in the 16 day rotation period case. 
\subsection{Detectability of molecular features}
\label{sec:dectability}
\begin{figure}
\centering
\includegraphics[width=0.5\textwidth]{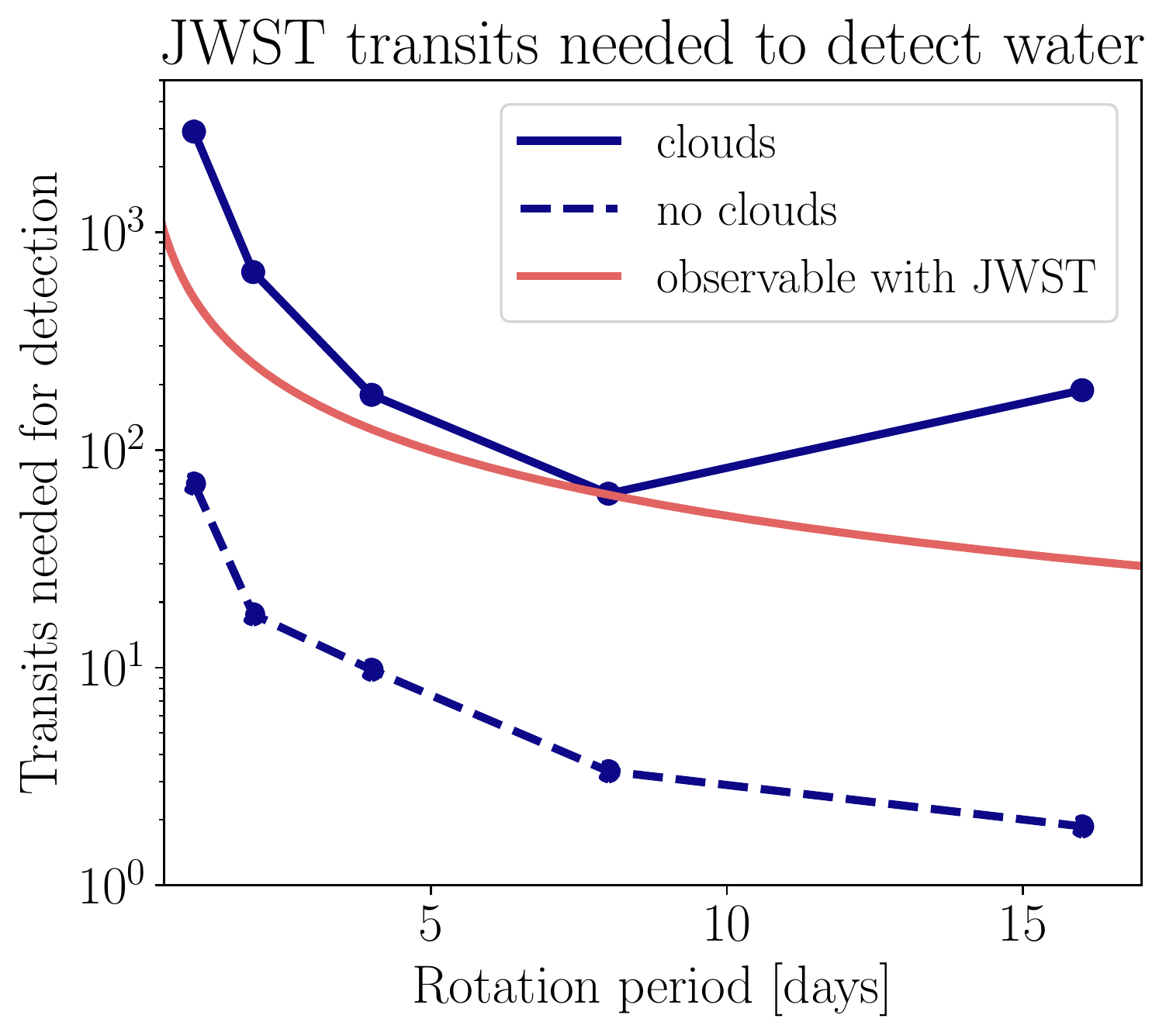}
\caption{\textbf{Clouds increase the number of transits needed to detect water vapor by one to two orders of magnitude}. Number of transits needed to detect water features both including and not including the effects of clouds from simulations with varying rotation period are shown in blue. The maximum number of transits observable with JWST for varying rotation period is shown in red, assuming the visibility of TRAPPIST-1 and synchronous rotation. Points show individual simulations for varying rotation period including clouds (solid lines connecting points) and not including clouds (dashed lines).}
%We assume an $\mathrm{SNR} \ge 5$ is required for detection.}
\label{fig:numtransit}
\end{figure}

\begin{table}
\begin{center}
\resizebox{0.5\textwidth}{!}{%
\begin{tabular}{| c | c | c |} 
\hline
Simulation parameters & Number of transits & Number of transits \\
 & with clouds & ignoring clouds \\
%\hline
\hline
{\bf Rotation period} & &  \\
%\hline
1 day & $>1000$ & 71 \\
2 days & 658 & 18  \\
4 days & 180 & 10  \\
8 days & 63 & 4  \\
16 days & 189 & 2 \\
\hline
{\bf Surface pressure} & &  \\
\textit{1 day rotation period:} & & \\
%\hline
%0.25 bars & $>1000$ & 3 (1) \\
0.5 bars & $>1000$ & 52  \\
1 bar & $>1000$  & 71  \\
2 bars & $>1000$ & 67   \\
4 bars & $>1000$ & 174 \\ 
\textit{8 day rotation period:} & & \\
0.5 bars & 24 & 2 \\
4 bars & 54 & 2 \\
\textit{16 day rotation period:} & & \\
0.5 bars & 68 & 1 \\
4 bars & 12 & 1 \\
\hline
{\bf Planetary radius} & &  \\
%\hline
\textit{1 day rotation period:} & & \\
0.5 $R_\varoplus$ & $>1000$ & 535 \\
0.707 $R_\varoplus$ & $>1000$ & 202 \\
1 $R_\varoplus$ &  $>1000$ & 71 \\
1.414 $R_\varoplus$ & $>1000$ & 32 \\
2 $R_\varoplus$ & 250 & 6 \\
\textit{8 day rotation period:} & & \\
0.5 $R_\varoplus$ & $>1000$ & 11 \\
2 $R_\varoplus$ & $>1000$ & 1 \\
\textit{16 day rotation period:} & & \\
0.5 $R_\varoplus$ & 904 & 5 \\
2 $R_\varoplus$ & 10 & 1 \\
\hline
{\bf Surface gravity} & & \\
%\hline
\textit{1 day rotation period:} & & \\
%0.5 $g_\varoplus$ & $>1000$ & 1 (1)  \\
0.707 $g_\varoplus$ & $>1000$ & 28 \\  
1 $g_\varoplus$ & $>1000$  & 71 \\  
1.414 $g_\varoplus$ & $>1000$ & 149 \\
\textit{8 day rotation period:} & & \\
0.707 $g_\varoplus$ & 479 & 1 \\
1.414 $g_\varoplus$ & 187 & 4 \\
\textit{16 day rotation period:} & & \\
0.707 $g_\varoplus$ & 27 & 1 \\
1.414 $g_\varoplus$ & 299 & 2 \\
\hline
{\bf Incident stellar flux} & &  \\
{\bf and rotation period} & &  \\
0.544 $F_\varoplus$, 6.49 days & 616 & 83 \\
0.667 $F_\varoplus$, 5.57 days & $>1000$ & 69 \\
0.816 $F_\varoplus$, 4.79 days & 709 & 43 \\
1 $F_\varoplus$, 4.11 days & 320 & 10 \\
\hline
\end{tabular}}
\caption{\textbf{Clouds dramatically increase the number of transits needed to detect water features.} Shown are the number of transits needed to detect water features at an $\mathrm{SNR} \ge 5$ with and without clouds for planets orbiting a late-type M dwarf star with $T_\mathrm{eff} = 2600 \ \mathrm{K}$ using JWST NIRSpec/Prism.  }
%Including our simulations for M-dwarf stars, this will result in a total of 43 simulations.}
\label{table:transit}
\end{center}
\end{table}
\indent Table \ref{table:transit} shows how the number of transits needed to detect water vapor with JWST NIRSpec/PRISM depends on planetary parameters, both including and not including the effect of clouds. We assume that a signal-to-noise ratio ($\mathrm{SNR}$) of 5 is required for detection. We determine the number of transits required to reach a given SNR using the method of \cite{Lustig-Yaeger:2019aa}, assuming that the SNR scales with the square root of the number of transit events. We perform this calculation for the maximum SNR of any water feature in the NIRSpec/PRISM bandpass, but note that the results are similar when considering the $1.4~\mu\mathrm{m}$ water feature (which does not overlap with a CO$_2$ feature) alone. \\
\indent Because TRAPPIST-1 is not continuously visible, \cite{Lustig-Yaeger:2019aa} find that JWST will have 123 opportunities in its nominal 5 year lifetime to observe the transit of TRAPPIST-1d, which has an orbital period of 4.05 days. Assuming the same visibility as TRAPPIST-1, the number of observable transits would decrease to 31 for a planet with an orbital period of 16 days. Note that this is the maximum observable number of transits, and a realistic JWST observing strategy would likely not capture every transit event. For our optimistic case of an Earth-sized planet with an 8 day rotation period, 62 transits are observable over the JWST lifetime, similar to the 63 needed to detect water vapor if its surface pressure is equal to that of Earth (see Table \ref{table:transit}).  \\
\indent We find that ignoring clouds, only $\sim 10$ transits would be required to detect water vapor in the atmosphere of a terrestrial exoplanet that orbits a late-type M dwarf star and receives an incident flux equal to that of Earth. The number of transits needed to detect water vapor decreases with increasing rotation period. This is because more slowly-rotating planets orbiting M dwarf stars have increased dayside convergence, leading to enhanced vertical transport of water vapor \citep{Komacek:2019aa}. When we include water clouds, the number of transits needed to detect water features is one to two orders of magnitude greater than when we do not include water clouds, for all considered variations in planetary parameters. We find that $63$ or more transits are required to detect water vapor in the atmospheres of Earth-sized planets with 1 bar atmospheres in the habitable zone around late-type M dwarf stars. \\
\indent \Fig{fig:numtransit} shows how the number of transits needed to detect water depends on rotation period in our suite of simulations. We find that the number of transits needed to detect water vapor sharply increases from $63$ to $189$ transits with increasing rotation period from 8 to 16 days. When not including clouds, $\lesssim 10$ transits are required to detect water vapor for planets with rotation periods $\ge 4~\mathrm{days}$. As a result, we expect that clouds will cause water vapor features to be challenging to detect in the atmospheres of Earth-sized planets with JWST. However, an extended mission lifetime, lowered SNR threshold for detection, or the discovery of a habitable planet that is continuously visible to JWST may allow for a detection of water vapor.
%table of transits needed for detection as a function of paraameters, cite lincowski paper for method
\subsection{Dependence of water vapor detection on planetary and atmospheric properties}
\indent Table \ref{table:transit} shows that the number of transits needed to detect water vapor is extremely sensitive to planetary parameters. We find that rotation period is a key controlling parameter. This is because planets that rotate quickly have reduced transport of water vapor to high altitudes, while planets that rotate slowly have significant high-altitude cloud cover at the terminator. When varying incident stellar flux and rotation period together, we find that there is a maximum in the number of transits needed to detect water vapor at an intermediate rotation period of 5.57 days. This is because there is a dynamical transition leading to decreased cloud cover for planets that are closer to their host star and more rapidly rotating \citep{Komacek:2019aa}, reducing the number of transits needed to detect water vapor. When varying surface pressure alone in the rapidly rotating regime and ignoring the effect of clouds, we find that the number of transits required to detect water vapor sharply increases with increasing surface pressure from 2 to 4 bars. This is because Rayleigh scattering from atmospheric N$_2$ increases with increasing surface pressure \citep{Kopp:2014}, leading to a cooler climate and reduced atmospheric water vapor content \citep{Komacek:2019aa}. \\
\indent In our simulations with a rotation period of 16 days, planets have ubiquitous high ice cloud coverage over a wide range of planetary parameters. Within this slowly-rotating regime, we find that planets that have larger radii, have lower gravities, and/or have higher surface pressures require fewer transits to detect water features. This is because, for fixed gravity, larger planets have a larger total transit signal, making deviations from the total signal larger. For fixed radius, lower-gravity planets have larger scale heights, leading to larger transmission features. Increasing the surface pressure leads to an increase in global-mean temperature and water vapor content of the air, leading to larger transmission features. The combination of large radius, high surface pressure, and low gravity enhances the detectability of water. Our results hence point toward sub-Neptunes (e.g., K2-18b, \citealp{Benneke:2019aa,Tsiaras:2019aa}) as viable targets to search for water vapor in transmission. \\
\indent In simulations with an intermediate rotation period of 8 days, we find that the cloud coverage is itself strongly affected by planetary parameters. The resulting trends in the number of transits needed to detect water are non-monotonic with increasing surface pressure, radius, and gravity. Higher surface pressure leads to a cloudier dayside \citep{Komacek:2019aa}, which increases the number of transits needed to detect water. However, at high surface pressures the increased amount of water vapor in the atmosphere makes it easier to detect. Larger radius leads to an increase in the size of transmission features, but at $2~R_\varoplus$ we find a transition to a more rapidly-rotating dynamical regime \citep{Yang:2019aa} that increases high-altitude ice cloud coverage and diminishes spectral features. Increased surface gravity makes the atmosphere clearer by causing the settling rate of cloud particles to increase. This is counteracted by the reduced atmospheric scale height of planets with larger gravity, making water hard to detect on high-gravity planets. 
\begin{table}
\begin{center}
%\resizebox{0.5\textwidth}{!}{%
\begin{tabular}{| c | c | c |} 
\hline
Ice cloud particle radius  & Number of transits  \\
%&  \\
%\hline
\hline
%{\bf Liquid cloud} &   \\
%{\bf particle radius} &  \\
%\hline
%7 $\mu\mathrm{m}$ & \\
%14 $\mu\mathrm{m}$ &  \\
%21 $\mu\mathrm{m}$ &  \\
%\hline
%{\bf Ice cloud} &   \\
%{\bf particle radius} &  \\
%\hline
20 $\mu\mathrm{m}$ & 126  \\
50 $\mu\mathrm{m}$ & 378  \\
80 $\mu\mathrm{m}$ & 377  \\
110 $\mu\mathrm{m}$ & 434  \\
140 $\mu\mathrm{m}$ & 327 \\
170 $\mu\mathrm{m}$ & 297 \\
200 $\mu\mathrm{m}$ & 222 \\
\hline
\end{tabular}
\caption{\textbf{Transmission spectral features of water are challenging to detect over a wide range of cloud particle size.} Shown are the number of transits needed to detect water features at an $\mathrm{SNR} \ge 5$ with ice clouds of varying particle size. These experiments have a rotation period of 16 days and all other parameters fixed to that of Earth.}
%Including our simulations for M-dwarf stars, this will result in a total of 43 simulations.}
\label{table:transitcld}
\end{center}
\end{table}
\\ \indent Our results are robust over a wide range in cloud particle size. Table \ref{table:transitcld} shows how the number of transits needed to detect water vapor on slowly-rotating terrestrial exoplanets orbiting late-type M dwarf stars depends on the ice cloud particle size. We find that for slowly-rotating planets with otherwise Earth-like planetary parameters, no cloud particle size we consider allows detection of water vapor in fewer than 100 transits, over a factor of ten variation in ice cloud particle size. Still, our results underly the importance of accurate microphysical modeling of cloud particle sizes, as plausible changes in ice cloud particle size can change the number of transits required for water detection by more than a factor of three. Additionally, we performed sensitivity tests including $400~\mathrm{ppm}$ of CO$_2$ and $1.7~\mathrm{ppm}$ of CH$_4$ in our GCM simulations. We found that the number of transits needed to detect water vapor when including CO$_2$ and CH$_4$ is comparable to or greater than that in simulations without CO$_2$ and CH$_4$.
%As a result, our results for the number of transits needed to detect water are optimistic, as including additional greenhouse gases makes water vapor more challenging to detect.
%reduce the amplitude of water features of habitable terrestrial planets significantly enough that they will be challenging to detect with JWST. 
%\indent \Fig{fig:numtransit} also shows that the number of transits needed to detect the N$_2$-N$_2$ CIA feature at $4.15 \ \mu\mathrm{m}$ strongly decreases with increasing surface pressure. For a surface pressure of 2 bars, we estimate that $194$ transits will be required to detect N$_2$-N$_2$ CIA, while for a surface pressure of 4 bars, only $31$ transits are required. As a result, in principle it may be possible to detect massive nitrogen atmospheres of terrestrial exoplanets orbiting M dwarf stars, as found by \cite{Lustig-Yaeger:2019aa}. However, we caution that in atmospheres with high amounts ($\gtrsim 0.5 \ \mathrm{bars}$) of CO$_2$, the N$_2$-N$_2$ feature can be overpowered by the wings of the $4.3 \ \mu\mathrm{m}$ CO$_2$ band \citep{Lincowski:2018aa}. The width of the $4.3 \ \mu\mathrm{m}$ CO$_2$ band increases in atmospheres with both high amounts of CO$_2$ and high background N$_2$, providing an additional probe of surface pressure \citep{Schwieterman:2015aa}.
%As a result, detailed modeling of the overlap in the shape of the N$_2$-N$_2$ and CO$_2$ features is required to determine whether the $4.15 \ \mu\mathrm{m}$ N$_2$-N$_2$ CIA feature would be detectable in a CO$_2$-rich atmosphere. 
\section{Discussion \& Conclusions}
\label{sec:disc}
\indent Our results are consistent with \cite{Lustig-Yaeger:2019aa,Fauchez:2019aa}, and \cite{Suissa:2019aa}, who also found that clouds will probably prevent the detection of water features on terrestrial planets via transit spectroscopy with JWST. 
%We find that for Earth-sized planets, no variation in planetary parameters that we considered changes our conclusion that water vapor features will be challenging to detect with JWST NIRSpec/PRISM. 
These results are also consistent with the non-detection of molecular features in the atmospheres of TRAPPIST-1d, e, and f with the Hubble Space Telescope \citep{Wit:2018aa}. However, water clouds only affect features originating from below the cloud deck, so well-mixed species that have strong spectral features (e.g., CO$_2$, CH$_4$) may still be detectable in the presence of water clouds \citep{Fauchez:2019aa}. As a result, searching for chemical disequilibrium as an exoplanet biosignature \citep{Kriss:2018aa,Kriss:2018ab} would likely not be significantly impacted by the presence of clouds. Similarly, a statistical search for variations in CO$_2$ as a function of position in the habitable zone might still be possible \citep{Bean:2017aa}. \\
%\indent Observations of emitted planetary radiation provide an avenue for studying the atmospheric composition of cloudy atmospheres, and can probe the spatial distribution of clouds through full-phase light curve measurements \citep{Yang:2013,Haqq2018,Wolf:2019aa}. For tidally locked terrestrial exoplanets orbiting late-type M dwarfs, secondary eclipse observations probe below the cloud layer because clouds are transported eastward of the substellar point due to the relatively fast planetary rotation \citep{Wolf:2019aa}. In this work, we also find downstream transport of clouds for fast-rotating planets orbiting late-type M dwarfs. As a result, it is possible that water in the atmospheres of slowly-rotating tidally locked planets will be detectable in emission with future space telescopes (e.g., OST). \\
%We will analyze the effects of varying planetary parameters on secondary eclipse emission spectroscopy and phase curve observations in future work. \\
\indent In this work, we did not consider atmospheres of planets that have significant amounts of water in the stratosphere or that are too hot to have surface liquid water. \cite{Fujii:2017aa} showed that atmospheres in a moist greenhouse state have strong water vapor spectral features. Further, \cite{Chen:2019aa} found that transmission spectral features of water vapor in these atmospheres could be detectable with JWST. Due to the increased scale height of runaway greenhouse atmospheres \citep{Turbet:2019aa}, observations of the atmospheres of terrestrial exoplanets orbiting M dwarfs that are interior to the habitable zone should find stronger molecular signatures. \\
%uncomment if long paper (e.g., \citealp{Wit:2016aa,Diamond-Lowe:2018aa}) should find stronger molecular signatures. \\
%Retrievals on direct imaging observations of terrestrial exoplanets with future missions (e.g., LUVOIR/HabEx) can provide constraints on the atmospheric water abundance of terrestrial exoplanets in the habitable zone \citep{Feng:2018aa}. \\
\indent Though we post-processed a complex GCM to simulate transmission spectra over a wide range of planetary parameters, there are a variety of limitations to our model setup. We did not consider all atmospheric constituents relevant for Earth-like planets, including O$_2$ and O$_3$. Additionally, we did not perform a retrieval on simulated spectra to quantify the effects of band overlap between molecular species on detectability. 
%We used an idealized model setup, varying one planetary parameter at a time. 
%As a result, when varying parameters other than rotation period, we analyze simulations from a fast rotating dynamical regime. However, our fast rotating simulations have relatively little dayside cloud cover, and as a result the effect of clouds on transmission spectra is weakened relative to the slowly rotating case. 
We did not include a dynamic ocean, which would affect the surface temperature distribution and location of dayside cloud cover \citep{Hu:2014aa,Genio:2017aa,way:2018,Yang:2019ab}. We did not include continents, which could reduce the amount of water vapor available to form clouds \citep{Lewis:2018aa}. Lastly, we assumed that water is plentiful on the surfaces of planets orbiting M dwarf stars. It is possible that planets orbiting M dwarf stars lose their entire surface complement of water, leading to high amounts of O$_2$ that could act as a false positive biosignature \citep{Ramirez2014,Luger2015,Tian2015,Schaefer:2016aa}.  \\
\indent In this work, 
%we simulated transmission spectra of tidally locked terrestrial exoplanets with a range of planetary parameters that orbit in the habitable zone of M dwarf stars. We 
we quantified the effect of water clouds on transmission spectra of tidally locked terrestrial exoplanets with a range of planetary parameters in the habitable zone of M dwarf stars. 
%by computing the number of transits required to detect water features both including and ignoring the effect of clouds. 
We find that transmission spectral features of water are significantly muted due to clouds on terrestrial exoplanets orbiting M dwarf stars. The decrease in transit depth due to clouds is especially strong for slowly rotating planets with rotation periods $\gtrsim 12 \ \mathrm{days}$, which have large dayside cloud decks. Due to cloud coverage, water transmission features of Earth-sized planets orbiting M dwarf stars will be challenging to detect with JWST. 
%For most planets, we find that water vapor is undetectable with JWST, with more than $1000$ transits required for detection.
%Transit features that originate from species that are mixed above the cloud deck (e.g., CO$_2$, O$_2$) are likely 
%\item Even when including clouds, high amounts of background N$_2$ may be detectable. N$_2$-N$_2$ collision-induced absorption is detectable in transmission for Earth-sized exoplanets orbiting M dwarf stars with atmospheric masses more than four times that of Earth and with little atmospheric CO$_2$. We find that $\lesssim 30$ transits are required to detect N$_2$-N$_2$ features in atmospheres with surface pressures $\gtrsim 4 \ \mathrm{bars}$. 
%\item We find that a certain amount of transits are required to see water features. With significant observing time investment (and if it's there), detection of water may be possible. 
%\end{enumerate}

\acknowledgements
We thank the referee for thoughtful comments that improved the manuscript. We acknowledge support from the NASA Astrobiology Institutes Virtual Planetary Laboratory, which is supported by NASA under cooperative agreement NNH05ZDA001C. T.D.K. acknowledges funding from the 51 Pegasi b Fellowship in Planetary Astronomy sponsored by the Heising-Simons Foundation. Our work was completed with resources provided by the University of Chicago Research Computing Center.

%\clearpage

\begin{thebibliography}{52}
\expandafter\ifx\csname natexlab\endcsname\relax\def\natexlab#1{#1}\fi

\bibitem[{Afrin Badhan {et~al.}(2019)Afrin Badhan, Wolf, Kopparapu, Arney, Kempton, Deming,
  \& Domagal-Goldman}]{Badhan:2019aa}
Afrin Badhan, M., Wolf, E., Kopparapu, R., Arney, G., Kempton, E., Deming, D., \&
  Domagal-Goldman, S. 2019, The Astrophysical Journal, 887, 34

\bibitem[{Barstow \& Irwin(2016)}]{Barstow:2016aa}
Barstow, J. \& Irwin, P. 2016, Monthly Notices of the Royal Astronomical
  Society, 461, L92
  
 \bibitem[{Batalha {et~al.}(2018)Batalha, Lewis, Line, Valenti, \& Stevenson}]{Batalha:2018aa}
Batalha, N.E., Lewis, N.K., Line, M.R., Valenti, J., \& Stevenson, K., 2018, The Astrophysical Journal Letters, 856, L34

\bibitem[{Bean {et~al.}(2017)Bean, Abbot, \& Kempton}]{Bean:2017aa}
Bean, J.L., Abbot, D.S., \& Kempton, E.M.-R., 2017, The Astrophysical Journal Letters, 841, L24

\bibitem[{Benneke {et~al.}(2019)Benneke, Wong, {et~al.}}]{Benneke:2019aa}
Benneke, B., Wong, I., Piaulet, C., Knutson, H.A., Lothringer, J., Morley, C.V., Crossfield, I.J.M., Gao, P., Greene, T.P., Dressing, C., Dragomir, D., Howard, A.W., McCullough, P.R., Kempton, E.M.-R., Fortney, J.J., \& Fraine, J. 2019, The Astrophysical Journal Letters, 887, L14
  
\bibitem[{Caldas {et~al.}(2019)Caldas, Leconte, Selsis, Waldmann, Borde,
  Rocchetto, \& Charnay}]{Caldas:2019aa}
Caldas, A., Leconte, J., Selsis, F., Waldmann, I., Borde, P., Rocchetto, M., \&
  Charnay, B. 2019, Astronomy {\&} Astrophysics, 623, A161

%\bibitem[{Carone {et~al.}(2015)Carone, Keppens, \& Decin}]{Carone:2015aa}
%Carone, L., Keppens, R., \& Decin, L. 2015, Monthly Notices of the Royal
  %Astronomical Society, 453, 2412

%\bibitem[{Charnay {et~al.}(2015)Charnay, Meadows, Misra, Leconte, \&
 % Arney}]{Charnay:2015}
%Charnay, B., Meadows, V., Misra, A., Leconte, J., \& Arney, G. 2015, The
%  Astrophysical Journal Letters, 813, L1
  
 \bibitem[{{Chen} {et~al.}(2019)}]{Chen:2019aa}
Chen, H., Wolf, E.T., Zhang, Z., \& Horton, D.E. 2019, arXiv e-prints:1907.10048

\bibitem[{Crossfield \& Kreidberg(2017)}]{Crossfield:2017aa}
Crossfield, I. \& Kreidberg, L. 2017, The Astronomical Journal, 154, 261

\bibitem[{de~Wit {et~al.}(2016)de~Wit, Wakeford, Gillon, Lewis, Valenti,
  Demory, Burgasser, Burdanov, Delrez, Jehin, Lederer, Queloz, Triaud, \&
  Grootel}]{Wit:2016aa}
de~Wit, J., Wakeford, H., Gillon, M., Lewis, N., Valenti, J., Demory, B.,
  Burgasser, A., Burdanov, A., Delrez, L., Jehin, E., Lederer, S., Queloz, D.,
  Triaud, A., \& Grootel, V.~V. 2016, Nature, 537, 69

\bibitem[{de~Wit {et~al.}(2018)de~Wit, Wakeford, Lewis, Delrez, Gillon, Selsi,
  Leconte, Demory, Bolmont, Bourrier, Burgasser, Grimm, Jehin, Lederer, Owen,
  Stamenkovi\'{c}, \& Triaud}]{Wit:2018aa}
de~Wit, J., Wakeford, H., Lewis, N., Delrez, L., Gillon, M., Selsi, F.,
  Leconte, J., Demory, B., Bolmont, E., Bourrier, V., Burgasser, A., Grimm, S.,
  Jehin, E., Lederer, S., Owen, J., Stamenkovi\'{c}, V., \& Triaud, A. 2018,
  Nature Astronomy, 2, 214

\bibitem[{{Del Genio} {et~al.}(2018){Del Genio}, Way, Amundsen, Aleinov,
  Kelley, Kiang, \& Clune}]{Genio:2017aa}
{Del Genio}, A., Way, M., Amundsen, D., Aleinov, I., Kelley, M., Kiang, N., \&
  Clune, T. 2018, Astrobiology, 19, 1

\bibitem[{Diamond-Lowe {et~al.}(2018)Diamond-Lowe, Berta-Thompson, Charbonneau,
  \& Kempton}]{Diamond-Lowe:2018aa}
Diamond-Lowe, H., Berta-Thompson, Z., Charbonneau, D., \& Kempton, E. 2018, The
  Astronomical Journal, 156, 42

\bibitem[{{Fauchez} {et~al.}(2019)}]{Fauchez:2019aa}
Fauchez, T.J., Turbet, M., Villanueva, G.L., Wolf, E.T., Arney, G., Kopparapu, R.K., Mandell, A., de Wit, J., Pidhorodetska, D., \& Domagal-Goldman, S.D. 2019, arXiv e-prints:1911.08596

\bibitem[{Feng {et~al.}(2018)Feng, Robinson, Fortney, Lupu, Marley, Lewis,
  Macintosh, \& Line}]{Feng:2018aa}
Feng, Y., Robinson, T., Fortney, J., Lupu, R., Marley, M., Lewis, N.,
  Macintosh, B., \& Line, M. 2018, The Astronomical Journal, 155, 200

\bibitem[{Fujii {et~al.}(2017)Fujii, Genio, \& Amundsen}]{Fujii:2017aa}
Fujii, Y., Genio, A.~D., \& Amundsen, D. 2017, The Astrophysical Journal, 848,
  100

\bibitem[{Gordon {et~al.}(2017)}]{Gordon:2017}
Gordon, I.E., Rothman, L.S., Hill, C., et al. 2017, Journal of Quantitative Spectroscopy and Radiative Transfer, 203, 3
  
\bibitem[{Greene {et~al.}(2016)Greene, Line, Montero, Fortney, Lustig-Yaeger,
  \& Luther}]{Greene:2015}
Greene, T., Line, M., Montero, C., Fortney, J., Lustig-Yaeger, J., \& Luther,
  K. 2016, The Astrophysical Journal, 817, 17

\bibitem[{Haqq-Misra {et~al.}(2018)Haqq-Misra, Wolf, Josh, Zhang, \&
  Kopparapu}]{Haqq2018}
Haqq-Misra, J., Wolf, E., Josh, M., Zhang, X., \& Kopparapu, R. 2018, The
  Astrophysical Journal, 852, 67

\bibitem[{Hu \& Yang(2014)}]{Hu:2014aa}
Hu, Y. \& Yang, J. 2014, PNAS, 111, 629

%\bibitem[{Joshi {et~al.}(1997)Joshi, Haberle, Reynolds}]{Joshi:1997}
%Joshi, M.M., Haberle, R.M., Reynolds, R.T. 1997, Icarus, 129, 450
  
 \bibitem[{Kasting {et~al.}(1993)Kasting, Whitmire, Reynolds}]{Kasting:1993aa}
Kasting, J.F., Whitmire, D.P., Reynolds, R.T. 1993, Icarus, 101, 108

 \bibitem[{Keller-Rudek {et~al.}(2013)}]{Keller-Rudek:2013aa}
Keller-Rudek, H., Moortgat, G.K., Sander, R., S\"{o}rensen, R., Earth System Science Data, 2013, 5, 365

%\bibitem[{Koll \& Abbot(2016)}]{Koll:2016}
%Koll, D. \& Abbot, D. 2016, The Astrophysical Journal, 825, 99

\bibitem[{Komacek \& Abbot(2019)}]{Komacek:2019aa}
Komacek, T. \& Abbot, D. 2019, The Astrophysical Journal, 871, 245
  
\bibitem[{Kopparapu {et~al.}(2014)}]{Kopp:2014}
Kopparapu, R.K., Ramirez, R.M., SchottelKotte, J., Kasting, J.F., Domagal-Goldman, S., \& Eymet, V. 2014, The Astrophysical Journal Letters, 787, L29

\bibitem[{Kopparapu {et~al.}(2016)}]{Kopp:2016}
Kopparapu, R.K., Wolf, E., Haqq-Misra, J., Yang, J., Kasting, J., Meadows, V.,
  Terrien, R., \& Mahadevan, S. 2016, The Astrophysical Journal, 819, 84

\bibitem[{Kopparapu {et~al.}(2017)Kopparapu, Wolf, Arney, Batalha, Haqq-Misra,
  Grimm, \& Heng}]{kopparapu2017}
Kopparapu, R.K., Wolf, E., Arney, G., Batalha, N., Haqq-Misra, J., Grimm, S., \&
  Heng, K. 2017, The Astrophysical Journal, 845, 5

\bibitem[{Kreidberg {et~al.}(2014)Kreidberg, Bean, D{\'{e}}sert, Benneke,
  Deming, Stevenson, Seager, Berta-Thompson, Seifahrt, \&
  Homeier}]{Kreidberg2014b}
Kreidberg, L., Bean, J.~L., D{\'{e}}sert, J.-M., Benneke, B., Deming, D.,
  Stevenson, K.~B., Seager, S., Berta-Thompson, Z., Seifahrt, A., \& Homeier,
  D. 2014, Nature, 505, 69

\bibitem[{Krissansen-Totton {et~al.}(2018a)Krissansen-Totton, Olson \& Catling}]{Kriss:2018aa}
Krissansen-Totton, J., Olson, S. \& Catling, D.C. 2018, Science Advances, 4, 5747

\bibitem[{Krissansen-Totton {et~al.}(2018b)Krissansen-Totton, Garland, Irwin \& Catling}]{Kriss:2018ab}
Krissansen-Totton, J., Garland, R., Irwin, P., \& Catling, D.C. 2018, The Astronomical Journal, 156, 114
  
\bibitem[{Lewis {et~al.}(2018)Lewis, Lambert, Boutle, Mayne, Manners, \&
  Acreman}]{Lewis:2018aa}
Lewis, N., Lambert, F., Boutle, I., Mayne, N., Manners, J., \& Acreman, D.
  2018, The Astrophysical Journal, 854, 171

\bibitem[{Lincowski {et~al.}(2019)Lincowski, Lustig-Yaeger, \&
  Meadows}]{Lincowski:2019aa}
Lincowski, A., Lustig-Yaeger, J., \& Meadows, V. 2019, The Astronomical
  Journal, 158, 26

\bibitem[{Lincowski {et~al.}(2018)Lincowski, Meadows, Crisp, Robinson, Luger,
  Lustig-Yaeger, \& Arney}]{Lincowski:2018aa}
Lincowski, A., Meadows, V., Crisp, D., Robinson, T., Luger, R., Lustig-Yaeger,
  J., \& Arney, G. 2018, The Astrophysical Journal, 867, 76

\bibitem[{Luger \& Barnes(2015)}]{Luger2015}
Luger, R. \& Barnes, R. 2015, Astrobiology, 15, 119

\bibitem[{Lustig-Yaeger {et~al.}(2019)Lustig-Yaeger, Meadows, \&
  Lincowski}]{Lustig-Yaeger:2019aa}
Lustig-Yaeger, J., Meadows, V., \& Lincowski, A. 2019, The Astronomical
  Journal, 158, 27

%\bibitem[{Merlis \& Schneider(2010)}]{Merlis:2010}
%Merlis, T. \& Schneider, T. 2010, Journal of Advances in Modeling Earth
 % Systems, 2, 13

\bibitem[{Moran {et~al.}(2018)Moran, H\"{o}rst, Batalha, Lewis, \& Wakeford}]{Moran:2018aa}
Moran, S.E., H\"{o}rst, S.M., Batalha, N.E., Lewis, N.K., \& Wakeford, H.R. 2018, The Astronomical Journal, 156, 252
  
\bibitem[{Morley {et~al.}(2017)Morley, Kreidberg, Rustamkulov, Robinson, \&
  Fortney}]{Morley:2017aa}
Morley, C., Kreidberg, L., Rustamkulov, Z., Robinson, T., \& Fortney, J. 2017,
  The Astrophysical Journal, 850, 121

%\bibitem[{Noda {et~al.}(2017)Noda, Ishiwatari, Nakajima, Takahashi, Takehiro,
 % Onishi, Hashimoto, Kuramoto, \& Hayashi}]{Noda:2017aa}
%Noda, S., Ishiwatari, M., Nakajima, K., Takahashi, Y., Takehiro, S., Onishi,
 % M., Hashimoto, G., Kuramoto, K., \& Hayashi, Y.-Y. 2017, Icarus, 282, 1

\bibitem[{Ramirez \& Kaltenegger(2014)}]{Ramirez2014}
Ramirez, R.M. \& Kaltenegger, L. 2014, The Astrophysical Journal Letters, 797, L25

\bibitem[{Rasch \& Kristj\'{a}nsson(1998)}]{Rasch:1998}
Rasch, P. \& Kristj\'{a}nsson, J. 1998, Journal of Climate, 11, 1587

\bibitem[{Schaefer {et~al.}(2016)Schaefer, Wordsworth, Berta-Thompson, \&
  Sasselov}]{Schaefer:2016aa}
Schaefer, L., Wordsworth, R., Berta-Thompson, Z., \& Sasselov, D. 2016, The
  Astrophysical Journal, 829, 63
  
\bibitem[{Sing {et~al.}(2015)Sing, Fortney, Nikolov, Wakeford, Kataria, Evans,
  Aigrain, Ballester, Burrows, Deming, D{\'{e}}sert, Gibson, Henry, Huitson,
  Knutson, des Etangs, Pont, Showman, Vidal-Madjar, Williamson, \&
  Wilson}]{Sing2015}
Sing, D.~K., Fortney, J.~J., Nikolov, N., Wakeford, H.~R., Kataria, T., Evans,
  T.~M., Aigrain, S., Ballester, G.~E., Burrows, A.~S., Deming, D.,
  D{\'{e}}sert, J.-M., Gibson, N.~P., Henry, G.~W., Huitson, C.~M., Knutson,
  H.~A., des Etangs, A.~L., Pont, F., Showman, A.~P., Vidal-Madjar, A.,
  Williamson, M.~H., \& Wilson, P.~A. 2015, Nature, 529, 18

\bibitem[{Suissa {et~al.}(2019)Suissa, Mandell, Wolf, Villanueva, Fauchez, \& Kopparapu}]{Suissa:2019aa}
Suissa, G., Mandell, A.M., Wolf, E.T., Villanueva, G.L., Fauchez, T., \& Kopparapu, R.K. 2019, Submitted
  
\bibitem[{Tian \& Ida(2015)}]{Tian2015}
Tian, F. \& Ida, S. 2015, Nature Geoscience Letters, 8, 5
  
%  \bibitem[{Turbet {et~al.}(2016)Turbet, Leconte, Selsis, Bolmont, Forget, Ribas,
 % Raymond, \& Anglada-Escud\'{e}}]{Turbet:2016aa}
%Turbet, M., Leconte, J., Selsis, F., Bolmont, E., Forget, F., Ribas, I.,
 % Raymond, S., \& Anglada-Escud\'{e}, G. 2016, Astronomy \& Astrophysics, 596,
  %A112
  
    \bibitem[{Tsiaras {et~al.}(2019)Tsiaras, Waldmann, Tinetti, Tennyson\& Yurchenko}]{Tsiaras:2019aa}
Tsiaras, A., Waldmann, I.P., Tinetti, G., Tennyson, J., \& Yurchenko, S.N. 2019, Nature Astronomy

  \bibitem[{Turbet {et~al.}(2019)Turbet, Enrenreich, Lovis, Bolmont, \& Fauchez}]{Turbet:2019aa}
Turbet, M., Ehrenreich, D., Lovis, C., Bolmont, E., \& Fauchez, T. 2019, Astronomy \& Astrophysics, 628, A12

\bibitem[{Villanueva {et~al.}(2018)Villanueva, Smith, Protopapa, Faggi, \&
  Mandell}]{Villanueva:2018aa}
Villanueva, G., Smith, M., Protopapa, S., Faggi, S., \& Mandell, A. 2018,
  Journal of Quantitative Spectroscopy and Radiative Transfer, 217, 86

\bibitem[{Way {et~al.}(2018)Way, Genio, Aleinov, Clune, Kelley, \&
  Kiang}]{way:2018}
Way, M., Genio, A.~D., Aleinov, I., Clune, T., Kelley, M., \& Kiang, N. 2018,
  The Astrophysical Journal Supplement Series, 239, 24

\bibitem[{Wolf(2017)}]{Wolf:2017aa}
Wolf, E. 2017, The Astrophysical Journal Letters, 839, L1

\bibitem[{Wolf {et~al.}(2019)Wolf, Kopparapu, \& Haqq-Misra}]{Wolf:2019aa}
Wolf, E., Kopparapu, R., \& Haqq-Misra, J. 2019, The Astrophysical Journal,
  877, 35

\bibitem[{Wolf {et~al.}(2017)Wolf, Shields, Kopparapu, Haqq-Misra, \&
  Toon}]{Wolf:2017}
Wolf, E., Shields, A., Kopparapu, R., Haqq-Misra, J., \& Toon, O. 2017, The
  Astrophysical Journal, 837, 107

\bibitem[{Wolf \& Toon(2015)}]{Wolf:2015}
Wolf, E. \& Toon, O. 2015, Journal of Geophysical Research: Atmospheres, 120,
  5775

\bibitem[{Yang {et~al.}(2019)Yang, Komacek, \&
  Abbot}]{Yang:2019aa}
Yang, H., Komacek, T., \& Abbot, D. 2019{\natexlab{a}}, The Astrophysical
  Journal Letters, 876, L27

\bibitem[{Yang {et~al.}(2019)Yang, Abbot, Koll, Hu, \&
  Showman}]{Yang:2019ab}
Yang, J., Abbot, D., Koll, D., Hu, Y., \& Showman, A. 2019{\natexlab{b}}, The
  Astrophysical Journal, 871, 29

\bibitem[{Yang {et~al.}(2014)Yang, Boue, Fabrycky, \& Abbot}]{Yang:2014}
Yang, J., Boue, G., Fabrycky, D., \& Abbot, D. 2014, The Astrophysical Journal
  Letters, 787, L2

\bibitem[{Yang {et~al.}(2013)Yang, Cowan, \& Abbot}]{Yang:2013}
Yang, J., Cowan, N., \& Abbot, D. 2013, The Astrophysical Journal Letters, 771,
  L45

\end{thebibliography}
\end{document}